%% file: charm2015_NahrgangMarlene.tex
\documentclass[12pt]{article}
\usepackage{graphicx}
\usepackage{amsmath,amssymb}
\usepackage{wrapfig}

\newcommand\pubnumber{WSU--HEP--XXYY}
\newcommand\pubdate{\today}

\def\wayne{Department of Physics\\
Duke University, Durham, NC 27708, USA}
\def\support{\footnote{Work supported by the postdoc programm of the German Academic Exchange Service (DAAD) and Department of Energy grant DE-FG02-05ER41367.}}

\def\Title#1{\begin{center} {\Large #1 } \end{center}}
\def\Author#1{\begin{center}{ \sc #1} \end{center}}
\def\Address#1{\begin{center}{ \it #1} \end{center}}

\newcommand\pubblock{\rightline{\begin{tabular}{l} \pubnumber\\
         \pubdate  \end{tabular}}}
\newenvironment{Abstract}{\begin{quotation}  }{\end{quotation}}
\newenvironment{Presented}{\begin{quotation} \begin{center} 
             PRESENTED AT\end{center}\bigskip 
      \begin{center}\begin{large}}{\end{large}\end{center} \end{quotation}}
\def\Acknowledgements{\bigskip  \bigskip \begin{center} \begin{large}
             \bf ACKNOWLEDGEMENTS \end{large}\end{center}}

\input econfmacros.tex

\begin{document}
\begin{titlepage}
\pubblock

\vfill
\Title{Open heavy-flavor production and suppression in heavy-ion collisions}
\vfill
\Author{Marlene Nahrgang\support}
\Address{\wayne}
\vfill
\begin{Abstract}
Heavy-flavor observables are valuable probes of the quark-gluon plasma, which is expected to be produced in ultrarelativistic heavy-ion collisions. These experiments offer the unique opportunity to study strongly interacting matter at high temperatures and densities in the laboratory. In this overview talk I will summarize the current theoretical status of heavy-flavor production and suppression in heavy-ion collisions and discuss open challenges.
\end{Abstract}
\vfill
\begin{Presented}
The 7th International Workshop on Charm Physics (CHARM 2015)\\
Detroit, MI, 18-22 May, 2015
\end{Presented}
\vfill
\end{titlepage}
\def\thefootnote{\fnsymbol{footnote}}
\setcounter{footnote}{0}
%

\section{Open heavy-flavor in heavy-ion collisions}
Ultrarelativistic heavy-ion collisions are currently performed at the LHC, CERN, and RHIC, BNL, with top beam energies of $\sqrt{s}_{\rm LHC} = 2.76 $~TeV (soon $5.1$~TeV) and $\sqrt{s}_{\rm RHIC} = 200 $~GeV. By analysing the final particle spectra we learn about the nature and the properties of the produced matter \cite{Jacak:2012dx}. The current theoretical and experimental status indicates the formation of the quark-gluon plasma (QGP) during the initial phase of the collision, which subsequently expands as an almost perfect fluid before hadronizing and rescattering in the final phase of hadronic interactions. The fluid dynamical behavior can be infered from measurements of flow harmonics, the Fourier coefficients $v_n$ of the azimuthal momentum anisotropy
\begin{equation}
 \frac{{\rm d}^3N}{p_T {\rm d} p_T {\rm d}\phi {\rm d} y}\propto 1+ 2\sum_n v_n \cos(n\phi)\, .
\end{equation}
This observable is very sensitive to the ratio of the shear viscosity over entropy density, $\eta/s$, of the QGP \cite{Song:2010mg,Schenke:2011bn}. 

Indications for the formation of the QGP come, for example, from jet modifications in heavy-ion collisions (AA) as compared to proton-proton collisions scaled by the number of binary collisions $N_{\rm coll}$ \cite{Bjorken:1982tu,Gyulassy:1990ye,Burke:2013yra}. This is expressed by the nuclear modification factor
\begin{equation}
 R_{\rm AA}(p_T)=\frac{1}{N_{\rm coll}}\frac{{\rm d}N_{AA}/{\rm d}p_T}{{\rm d}N_{pp}/{\rm d}p_T}\, .
\end{equation}

While the flow measurements are interesting for the properties of the bulk, the nuclear modification factor gives insight into the individual scattering processes of probe particles at high transverse momentum and is sensitive to the jet-quenching parameter $\hat{q}$.

The mass of heavy quarks ($m_c\sim 1.3-1.5$~GeV and $m_b\sim 4.5-5.1$~GeV) sets another scale and lets us investigate both regimes: for low momenta ($p_T\le m_{\rm HQ}$) the heavy-quark diffusion occurs mainly via collisional processes and can lead to the partial thermalization with the QGP medium. Here, additional effects like nuclear shadowing in the initial state and hadronization via recombination can affect the results. At high momenta ($p_T\ge m_{\rm HQ}$) perturbative QCD calculations should be applicable and the heavy quarks are expected to behave similarly to the light partons.

The investigation of heavy-quark production and suppression in heavy-ion collisions has a couple of advantages: the heavy-flavor vacuum shower terminates earlier than for the light partons because of the reduction of the formation time by the heavy-quark mass: $\tau_f=1/Q_{\rm HQ}=1/\sqrt{Q_0^2+m_{\rm HQ}^2}$. This means the heavy quark is on-shell before the surrounding QGP medium has formed ($\tau_{\rm QGP}\sim0.3-1.0$~fm/c). During the QGP expansion the temperature is well below the heavy-quark mass such that the number of thermally excited heavy quarks is small.
During the evolution in the QGP medium a heavy quark as leading parton remains tagged. The emission of a hard gluon only changes the energy of the leading parton, not its identity. 

This overview is organized as follows. In section \ref{sec:transport}, I will discuss the various steps of describing heavy-flavor transport from production and evolution in the medium to hadronization. This includes the evolution of the medium itself and its effect on the heavy-flavor observables. A brief comparison to experimental observables will be given in section \ref{sec:expobs} before a summarizing open challenges in section \ref{sec:summary}.

\section{Heavy-flavor transport: from production to the detector}\label{sec:transport}
The description of heavy-flavor transport in heavy-ion collisions proceeds in various steps. During the initial hard scatterings the heavy-quarks are produced. In this stage the QGP forms and upon the early thermalization time the heavy quarks interact with the QGP constituents. As the medium expands and dilutes heavy-flavor mesons and baryons are formed. Although a small cross section is expected the final interactions of heavy-flavor hadrons with the hadronic medium might affect the results as well.

\subsection{Production}
The momentum and rapidity spectra for the initial production of heavy quarks can be obtained from LO pQCD calculations. As such, however, they do not describe the experimental data satisfactorily. Fixed-order and next-to-leading log resummation, FONLL, give inclusive spectra which are in excellent agreement with the available data \cite{FONLL}. Because all additional degrees of freedom are integrated out, the resulting spectra have no information about $Q\bar{Q}$ pair production. The state-of-the-art methods for more exclusive spectra, like azimuthal $Q\bar{Q}$ correlations, are event generators, which couple next-to-leading order pQCD matrix elements to parton shower, like POWHEG \cite{Frixione1} or MC@NLO \cite{Frixione2}. These approaches work well for the pair production of bottom quarks, but a substantial discreptancy to the measured $D\bar{D}$ correlations in proton-proton collisions are observed \cite{CMSbbbar}. Here, further theoretical and experimental effort is needed, as two-particle correlations turn out to be promising new observables, as will be discussed in the final section of this overview.

Embedding the production spectra in elementary proton-proton collisions in the initial state of heavy-ion collisions, one needs to consistently couple the soft sector of light partons and the hard production processes in both momentum and coordinate space. Currently, most theoretical models do not fully capture the relation between the underlying event activity and the hard production cross section \cite{Adam:2015ota}.

\subsection{Interaction with QGP constituents}
The best studied transport equation in physics is probably the Boltzmann equation, which describes the evolution of the phase-space distribution of the particles of interest, here the heavy quarks,
\begin{align}
 \frac{{\rm d}}{{\rm d}t}f_{\rm HQ}(t,\vec{x},\vec{p})=&{\cal C}[f_{\rm HQ}]\\
 {\rm with}\quad {\cal C}[f_{HQ}]=&\int{\rm d}\vec{k}[\underbrace{w(\vec{p}+\vec{k},\vec{k})f_{\rm HQ}(\vec{p}+\vec{k})}_{\rm  gain\, term}-\underbrace{w(\vec{p},\vec{k})f_{\rm HQ}(\vec{p})}_{\rm  loss\, term}]\, ,
\end{align}
where $w(\vec{p},\vec{k})$ is the transition rate for $\vec{p}\to\vec{p}-\vec{k}$. Then the first contribution in the collision integral ${\cal C}$ is a gain term and the second contribution is the loss term. If one 
expands ${\cal C}$ for small momentum transfer $k\ll p$, where a typical momentum transfer in the thermal medium is $k\sim{\cal O}(gT)$, and keeps only the lowest two terms, one arrives at the Fokker-Planck equation
\begin{equation} 
\frac{\partial}{\partial t}f_Q(t,\vec{p})=\frac{\partial}{\partial{p^i}}\left( {A^i(\vec{p})}f_Q(t,\vec{p})+\frac{\partial}{\partial{p^j}}\left[{B^{ij}(\vec{p})}f_Q(t,\vec{p})\right]\right)\, .
\end{equation}
Here, $A^i(\vec{p})$ is the drag coefficient describing the friction of the heavy quark in the medium and $B^{ij}(\vec{p})$ is the momentum diffusion. One can recast the Fokker-Planck equation to the Langevin equation, which describes the stochastic trajectories of a Brownian motion particle.
\begin{equation}
 \frac{{\rm d}}{{\rm d}t}\vec{p}=-\eta_D(p)\vec{p}+\vec{\xi}\quad{\rm with}\quad \langle\xi^i(t)\xi^j(t')\rangle=\kappa^{ij}(\vec{p})\delta(t-t')
\end{equation}
and $\kappa^{ij}(\vec{p})=\kappa_L(p)\hat{p}^i\hat{p}^j+\kappa_T(p)(\delta^{ij}-\hat{p}^i\hat{p}^j)$. Often the approximation $\kappa_L=\kappa_T=\kappa$ is applied. Then the transport coefficients can be connected by the fluctuation-dissipation theorem (Einstein relation)
$\eta_D=\kappa/2m_{\rm HQ}T$
in order to achieve thermalization to the correct equilibrium state in the long-time limit. The spatial diffusion coefficient is related to the drag coefficient via $D_s=T/m_{\rm HQ}\eta_D$.

While the Fokker-Planck/Langevin approximation is probably valid for the heavier bottom quark, differences to the description via the full Boltzmann equation become significant for the charm quarks \cite{Das:2013kea}. Most approaches to heavy-quark propagation still use the Langevin equation, which is easy to implement numerically \cite{vanHeesRapp,langevin}, while some apply the full Boltzmann equation either in a parton cascade \cite{bamps} or via thermal sampling of the medium constituents \cite{Gossiaux:2008jv,Nahrgang:2013xaa}. The Boltzmann equation itself is applicable in a dilute medium, where the incoming particles are asymptotic on-shell states undergoing independent scatterings. Recently, also off-shell transport via the Kadanoff-Baym equation was applied to study $D$ meson dynamics \cite{Song:2015sfa}.

\begin{wrapfigure}{r}{0.5\textwidth}
  \centering
 \includegraphics[width=0.5\textwidth]{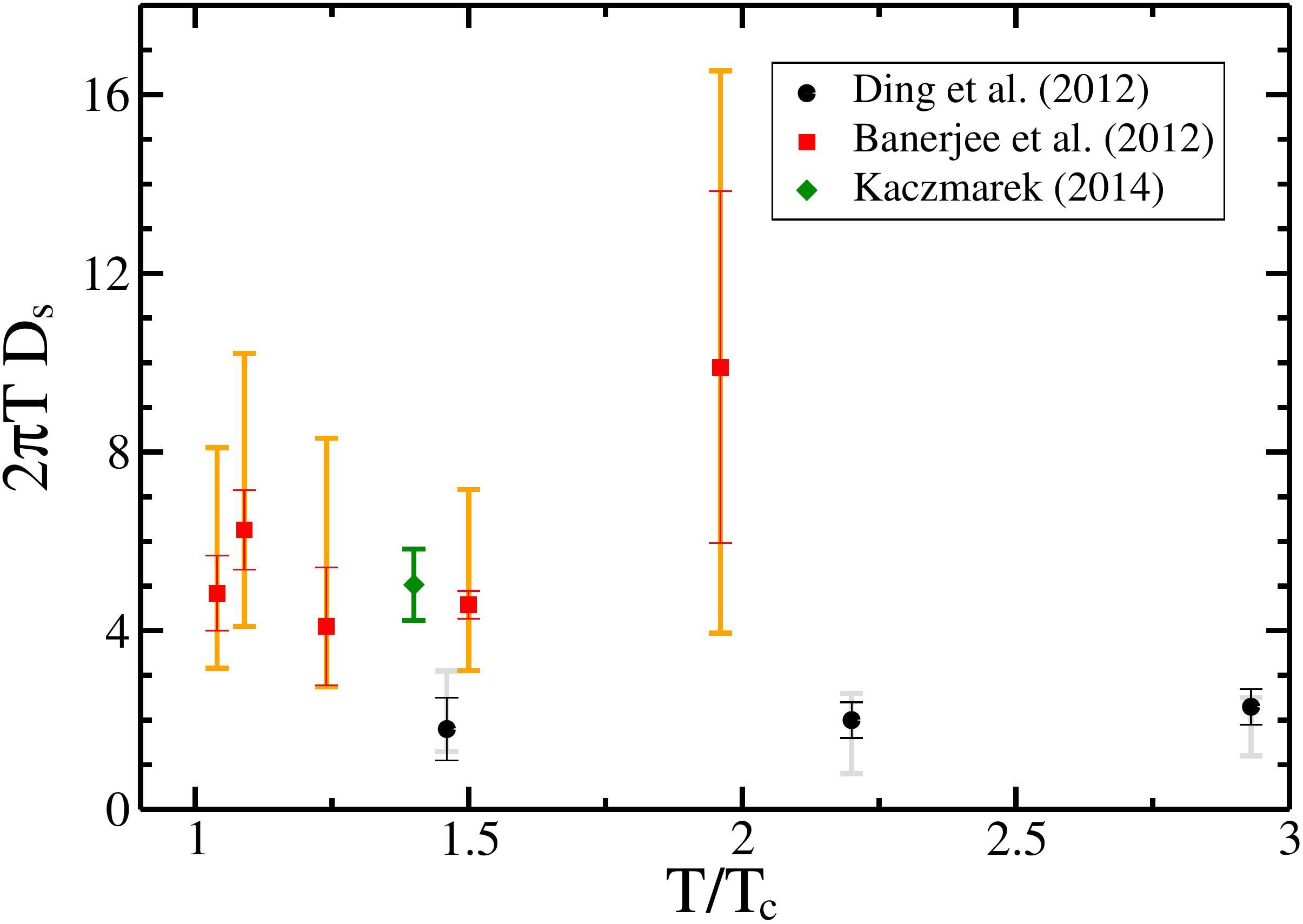}
  \caption{The spatial diffusion coefficient from two different lattice calculations on finite lattices and the continuum extrapolated value by Kaczmarek \cite{lattice}.}
  \label{fig:lQCD}
\end{wrapfigure}

Regardless of the particular equation used, transport coefficients or cross sections need to be obtained from the underlying theory. For temperatures close to $\Lambda_{\rm QCD}$ and thus inside the nonperturbative regime, one would naturally turn to lattice QCD calculations. Lattice QCD at finite $T$ is, however, performed in Euclidean space and it is therefore notoriously difficult to calculate dynamical quantities. Transport coefficients need thus be infered from calculable quantities on the lattice, such as from the correlation function of conserved currents via the slope of the spectral function $\rho_E$ at $\omega=0$ according to the Kubo formula. Then the momentum diffusion is obtained from $\kappa/T^3=\lim_{\omega\to0} 2T\rho_E(\omega)/\omega$. As can be seen from Fig.~\ref{fig:lQCD} the current uncertainty from lattice QCD calculations is still large and cannot be used as a reliable input.

\subsubsection{Perturbative QCD calculations: elastic cross sections}
In the leading order Feynmann diagrams for perturbative heavy-quark scattering off a light parton the $t$-channel diagram gives the dominant contribution. It requires, however, an infrared regularisation, schematically taken as the Debye screening mass $m_{\rm D}\sim{\cal O}(gT)$. This calculation can be extended by the hard-thermal loop (HTL) resummation \cite{Moore:2004tg}. Here, the bare gluon propagator is regularized by the HTL gluon propagator for small momentum transfers $|t|\ll t^*$. For well-separated scales $g^2T^2\ll T^2$ the average energy loss becomes independent of the intermediate scale $t^*$ \cite{Braaten}. In the temperature regime reached in heavy-ion collisions this is most likely not the case. In the model proposed by the Nantes group \cite{Gossiaux:2008jv} a reduced IR regulator $\lambda m_{\rm D}^2$ is included in the hard part of the propagator (semi-hard), with $\lambda$ determined such as to achieve maximal independence of the intermediate scale $t^*$. Finally, a scalar one-gluon exchange propagator $G(t)=\frac{\alpha_s}{t-\kappa \tilde m_{\rm D}^2}$  is used over the whole momentum range $t$, where $\kappa$ is tuned such that the average energy loss calculated in the HTL+semi-hard model is reproduced. A running coupling constant $\alpha_s(Q^2)$ is applied \cite{Dokshitzer:1995qm} and the Debye-mass is evaluated self-consistently as $\tilde m_{\rm D}^2=(1+6n_f)4\pi\alpha_s(\tilde m_{\rm D}^2) T^2$ \cite{Peshier:2006hi}. The transition matrix elements are then calculated in the Born approximation. 

\subsubsection{Perturbative QCD calculations: radiative processes}
In the ultra-relativistic regime, $p_{\rm HQ}\gg m_{\rm HQ}$, the emission of gluons as a result of the scattering of a heavy quark inside the medium is usually considered to be primary source of energy loss. In this regime the formation time of gluons is long and several scatterings lead to a coherent emission, such that the total energy loss can only be calculated over the total path length of the heavy quark in the medium. This effect of coherent radiation is similar to the Landau-Pomeranchuk-Migdal effect in QED, with the important difference that the emitted gluon can further decohere by the interaction with medium gluons  \cite{BDMPSZ}. The suppression of radiation is thus smaller in QCD than in QED. A dynamical realization of the coherent emission is challenging \cite{Zapp} and most approaches only consider an effective suppression of the emission spectra. At relativistic energies this coherence effect is less important and the radiation can be described by several incoherent scatterings. The leading-order pQCD matrix element for $2\to3$ process have been calculated in \cite{Kunszt:1980} 
and the Gunion-Bertsch approximation \cite{Gunion:1981qs} is derived in the high-energy limit, where the $k_\perp$ of the radiated gluon and the momentum transfer $q_\perp$ are soft $\ll\sqrt{s}$. In \cite{Fochler:2013epa,Aichelin:2013mra} the distribution of induced gluon radiation has been extended to finite heavy-quark masses. The obtained forms can readily be implemented in Monte-Carlo simulations of the Boltzmann equation taking exact momentum conservation and scattering on dynamical medium constituents into account. 
A description of average radiative energy loss including many important effects, such as dynamical medium constituents, finite size effects, production inside the medium, finite magnetic mass and running coupling, has been developed in \cite{Djordjevic}. Calculations of the nuclear modification factor, however, are performed at a fixed temperature without an explicit evolution of the QGP medium. 

\subsection{Evolution of the medium}\label{sec:evolmedium}
In order to describe the heavy-quark dynamics inside the expanding QGP medium a model for the partonic phase needs to provide the scattering partners for the heavy quarks. This can either be a microscopic transport model in terms of a parton cascade \cite{bamps} or a fluid dynamical description. In the latter case, the scattering partners will be sampled from a thermal distribution, which takes as inputs the local temperature and fluid velocity fields. For a reliable quantitative comparison to experimental data it is of course highly important to use a medium description that reproduces well the bulk observables of light hadrons.

\subsection{Hadronization and final-state interactions}
When the system becomes dilute the heavy quarks eventually hadronize together with the bulk medium. This is typically described on a hypersurface of constant temperature or energy density. There are two competing hadronization mechanism involved, at predominantly low $p_T$ the heavy quarks are assumed to coalesce \cite{reco} with a quark from the medium to form $D$ or $B$ mesons, while at high $p_T$ the heavy quarks predominantly fragment \cite{Kartvelishvili:1977pi}. In the first case, the $D$ or $B$ meson will have a larger $p_T$ than the heavy quark, while during the fragmentation process the heavy quark loses some of its $p_T$ due to the emission of gluons. The hadronic interaction of $D$ and $B$ mesons was estimated to be small, but an inclusion of final-state hadronic interaction can still contribute $10$-$20$\% depending on the observable \cite{hadron}.

\section{Comparison to experimental observables}\label{sec:expobs}
\begin{figure}
 \centering
 \includegraphics[width=0.45\textwidth]{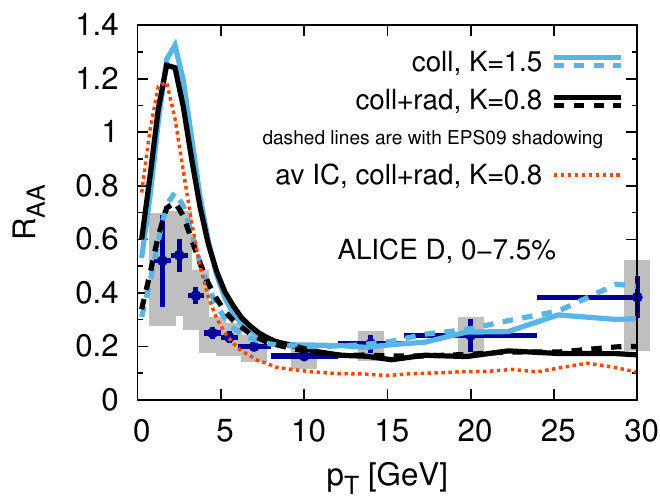}\hfil
 \includegraphics[width=0.45\textwidth]{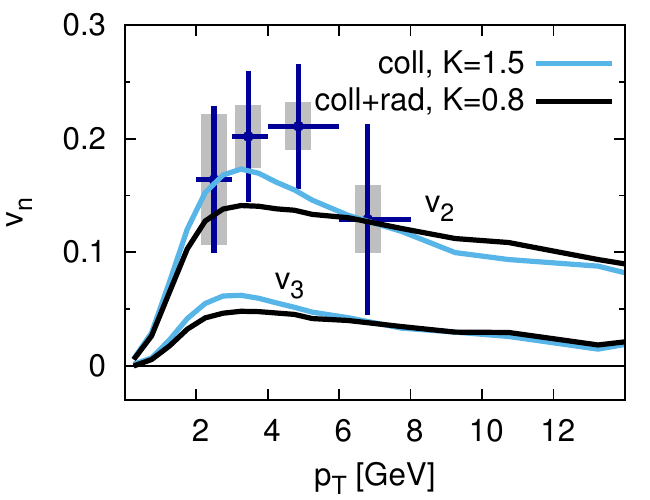}
  \caption{The $R_{\rm AA}$, $v_2$ and $v_3$ calculated from the MC@sHQ+EPOS2 model for Pb+Pb collisions at $\sqrt{s}=2.76$~TeV compared to data from ALICE \cite{ALICE}.}
  \label{fig:mcatshq}
\end{figure}

In Fig.~\ref{fig:mcatshq} we exemplarily show results from the MC@sHQ+EPOS2 model \cite{Nahrgang:2013xaa,Nahrgang:2013saa, Nahrgang:2014vza} based on a coupling between MC@sHQ \cite{Gossiaux:2008jv} and the fluid dynamical evolution from the EPOS event generator \cite{EPOS}. The overall agreement with the experimental data is very good for both observables. For the calculation of the $R_{\rm AA}$ the purely collisional and the collisional plus radiative, including an effective suppression due to the coherent radiation, are compared. A global scaling of the cross sections is performed by a $K$-factor. Including the radiative energy loss one finds a $K$-factor close to unity. In the low $p_T$ regime the inclusion of shadowing, here via the EPS09 parametrization of the nuclear parton distribution function \cite{Eskola:2009uj}, is necessary to reproduce the data. At high $p_T$ the rising trend of the purely collisional energy loss is slightly favored by the data. The standard calculations are performed with event-by-event initial conditions, but we compare here to a calculation performed with smooth, averaged initial conditions. As discussed in \cite{Nahrgang:2014mla} the smooth initial conditions lead to an enhanced energy loss. This underlines the importance of using most realistic descriptions of the evolution of the QGP and gives an idea of the uncertainty which is generally involved in comparing models to data. 
A detailed overview over many theoretical models and their comparison to the available experimental data is given in \cite{Andronic:2015wma}.

 
\section{Challenges and Outlook}\label{sec:summary}
The qualitative features of heavy-quark dynamics as reflected in the nuclear modification factor and the elliptic flow can nowadays be described by a large variety of models based on pQCD-inspired interactions, nonperturbative scatterings \cite{vanHeesRapp} or strong-coupling approaches \cite{Horowitz:2015dta}. Given the large experimental error bars on the currently available data from RHIC and LHC experiments the discriminative power, especially for the model ingredients determining the transport coefficients, is limited.

With the upcoming high-statistics data from the next runs at RHIC, including the heavy-flavor tracker at the STAR experiment, and at the LHC one hopes to be able to better constrain the heavy-quark dynamics in the QGP. Another approach is to study different systems, such as Cu+Cu or p+Pb collisions. Especially the asymmetric, small system shall be important to distinguish between initial state cold nuclear matter effects and interactions inside the hot medium. It is currently under debate if the QGP is formed in p+Pb collision and whether the measured light hadron flow harmonics are indicative for a fluid dynamical phase. At present the minimum bias $R_{pPb}$ for $D$ mesons does not show a suppression at higher $p_T$ within large error bars and is consistent with purely initial state effects. It remains a very interesting question if events with higher multiplicity show final state suppression as well and how models are able to reproduce this.

Besides different systems, the new data will hopefully allow us to investigate other observables than the traditional ones. In the light-hadron sector the success of pinning down the ratio of shear viscosity over entropy density has been significantly supported by the measurements of correlations and higher-order flow harmonics. Recent studies indicate that these observables shall as well provide additional information about the heavy-quark interaction with the medium as well. The azimuthal correlations between $c\bar{c}$ or $b\bar{b}$ pairs are very sensitive to the contributions of purely elastic scatterings versus gluon emission \cite{Nahrgang:2013saa,Cao:2015cba}. At low momenta the initially correlated $c\bar{c}$ pairs decorrelate almost completely and their participation in the collective flow of the medium results in the typical $v_2$-like two-particle correlation pattern, in case of purely elastic interactions. The inclusion of gluon emission leads to substantially less broadening of the azimuthal angle. It remains unclear if experimentally it will be possible to measure $D\bar{D}$ correlations or correlations between $D$ mesons and heavy-flavor decay electrons directly or if only $D$-hadron correlations will be accessible. In order to describe $D$-hadron correlations current models need to include more aspects of the coupling of heavy to light-flavor observables.
In addition to correlations, higher-order flow harmonics are predicted to show the incomplete coupling of the $D$ and $B$ mesons to the low-momentum bulk flow more clearly and could thus play an important role in determining the diffusion coefficient as well as to understand the processes that lead to partial thermalization of the charm quarks inside the medium \cite{Nahrgang:2014vza}. 

\Acknowledgements
M.N. acknowledges support from a fellowship within the Postdoc-Program of the German Academic Exchange Service (DAAD). This work was supported by the U.S. department of Energy under grant DE-FG02-05ER41367.


\end{document}

%% file: econfmacros.tex
\def\beq{\begin{equation}}
\def\eeq#1{\label{#1}\end{equation}}
\def\eeqn{\end{equation}}

\def\beqa{\begin{eqnarray}}
\def\eeqa#1{\label{#1}\end{eqnarray}}
\def\eeqan{\end{eqnarray}}

\let\bar=\overbar

\def\Dslash{\not{\hbox{\kern-4pt $D$}}}
\def\dslash{\not{\hbox{\kern-2pt $\del$}}}

\def\msb{{\bar{\ssstyle M \kern -1pt S}}}

